# CMOS+X: Stacking Persistent Embedded Memories based on Oxide Transistors upon GPGPU Platforms

Faaiq Waqar†, Ming-Yen Lee, Seongwon Yoon, Seongkwang Lim, Shimeng Yu†
School of Electrical and Computer Engineering, Georgia Institute of Technology, Atlanta, GA, USA
†email: faaiq.waqar@gatech.edu, shimeng.yu@ece.gatech.edu

## ABSTRACT

In contemporary general-purpose graphics processing units (GPGPUs), the continued increase in raw arithmetic throughput is constrained by the capabilities of the register file (single-cycle) and last-level cache (high bandwidth), which require the delivery of operands at a cadence demanded by wide single-instruction multiple-data (SIMD) lanes. Enhancing the capacity, density, or bandwidth of these memories can unlock substantial performance gains; however, the recent stagnation of SRAM bit-cell scaling leads to inequivalent losses in compute density.

To address the challenges posed by SRAM's scaling and leakage power consumption, this paper explores the potential CMOS+X integration of amorphous oxide semiconductor (AOS) transistors in capacitive, persistent memory topologies (e.g., 1T1C eDRAM, 2T0C/3T0C Gain Cell) as alternative cells in multi-ported and high-bandwidth banked GPGPU memories. A detailed study of the density and energy tradeoffs of back-end-of-line (BEOL) integrated memories utilizing monolithic 3D (M3D)-integrated multiplexed arrays is conducted, while accounting for the macro-level limitations of integrating AOS candidate structures proposed by the device community —an aspect often overlooked in prior work. By exploiting the short lifetime of register operands, we propose a multi-ported AOS gain-cell capable of delivering 3× the read ports in ~76% of the footprint of SRAM with >70% lower standby power, enabling enhancements to compute capacity, such as larger warp sizes or processor counts. Benchmarks run on a validated NVIDIA Ampere-class GPU model, using a modified version of Accel-Sim, demonstrate improvements of up to 5.2× the performance per watt and an average 8% higher geometric mean instruction per cycle (IPC) on various compute- and memory-bound tasks.

## 1. Introduction

Graphics processing units (GPUs) have become pivotal to server scaling, as noted by NVIDIA's 427% increase in data-center revenue from FY24 to FY25 [1]. Though seminal engineering efforts of the GPU targeted the acceleration of raster graphics and video workloads [2] (as their namesake implies), their combination of high-bandwidth Single-Instruction-Multiple-Data (SIMD) execution units (stream multiprocessors, SMs) and on-chip memories have proved ideal for highly parallelized processing of numerically intensive floating-point arithmetic [3], leading to their ubiquity in the processing of modern AI/ML training and inference and large-scale scientific computing.

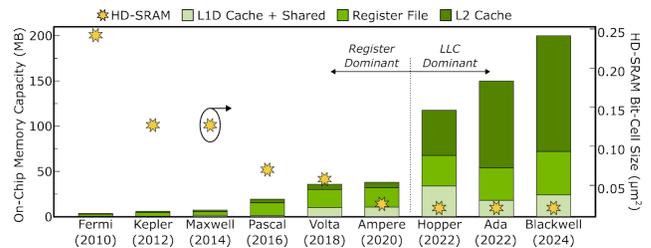

**Fig 1:** Scaling of GPU shared memory, registers, L1, and last level cache (LLC) vs corresponding SRAM cell in NVIDIA architectures

The performance of a GPU is often bottlenecked by the available bandwidth, capacity, and die area of on-chip memory subsystems [4]. In workloads with low arithmetic intensity, SMs frequently idle while awaiting memory accesses during execution, resulting in underutilization of execution units [5]. Scaling the SM count intensifies contention for the shared L2 cache, while the number of resident warps per SM is capped by the size of its register file (RF) and the per-thread register allocation. Raising the warps-per-SM budget within a fixed die area, therefore, trades off against the total number of SMs that can be integrated [6]. Since modern GPUs support many registers per thread, the compiler directs operand spillover into the L1 data cache (L1D) when register file capacity is exhausted, imposing increased traffic and occupancy pressure on the slower L1D [7]. Tasks such as backpropagation and blackscholes, which demonstrate low intra-warp divergence, benefit from higher threads per warp; however, the number of operands demanded per cycle (and hence the required ports or banks) scales linearly with the warp size [8]. The imposition of these memory requirements has inevitably led to the sharp rise in cumulative memory capacity (especially that of the register files and the L2) in modern GPUs by over two orders of magnitude from NVIDIA's Fermi (2010) to Blackwell (2024) architectures [9], while the density of the high density SRAM

(HD-SRAM) bit cell has only climbed by one order of magnitude (Fig. 1). This disparity is the key indication that **GPU memory scaling is driven by architectural demand, not fundamental SRAM densification**, and as such, is limited by SRAM technology scaling constraints.

Recent advances in semiconductor fabrication have enabled the integration of active devices, such as non-volatile memories, in the back-end-of-line (BEOL) stack. By relying on low-temperature (<400 °C) deposition steps, multiple tiers of memory can be fabricated above the front-end-of-line (FEOL) logic without degrading the underlying CMOS transistors [10]. Among the most promising options are amorphous-oxide-semiconductor (AOS) transistors, which combine ultra-low off-state leakage (< fA/μm; three to four orders of magnitude lower than Si MOSFETs) with adequate electron mobility (~ 20 cm$^2$/V·s) [11]. Leading research institutes, such as IMEC, have demonstrated functional prototypes based on AOS 2T0C arrays [12], and major foundries, including TSMC, have also reported AOS 1T1C macros with high yield [13]. Prior modelling studies show that AOS 2T0C memories can outperform SRAM in TPU buffers [14], boost energy efficiency in digital compute-in-memory (DCIM) accelerators [15], [16] and deliver ~4.5 × higher density (at 256 MB) with nominal performance improvements when deployed as a shared CPU L3 cache [17]. We hypothesize that, in gain-cell configurations (§4.2, §5) with small storage node capacitance and decoupled read/write paths, the speed of AOS gain cell memories may be sufficient to serve as single-cycle register memories in GPUs with *lower base clock frequency* than their CPU counterparts [18]. Moreover, to meet the needs of memory-bound tasks, several AOS-based candidates, such as BEOL-compatible 1T1C eDRAM and 2T0C/3T0C gain cell topologies, offer a practical avenue to enhance the overall bandwidth and density of the large shared L2. To understand the viability of these respective memory candidates, a critical evaluation of their achievable density, bandwidth, and energy efficiency is required under constrained design exploration (e.g., imposed by sneak-path currents, increased parasitic capacitance, IR drop, lower mobility), which must be well-characterized. In performing this, this paper makes the following contributions:

• Using Accel-Sim [19], we evaluate the lifetime of operands in GPU register files to determine the retention requirements of register memories. Furthermore, we propose a stacked multi-read port AOS gain-cell (NT0C) that enables 3× the read ports in ~76% of the footprint of a comparative 8T-SRAM bank.

• Using NS-cache [17], we analyze the limitations in AOS 2T0C array scaling under sneak path and IR drop constraints studied in SPICE, 1T1C access time and sense margin tradeoffs, and 3T0C gate loading and leakage under read path threshold voltage. We demonstrate that (1) the $f_{max}$ limitations on peak bandwidth of AOS 1T1C banks are easily overcome through increased partitioning, (2) the AOS 3T0C speed vs. leakage tradeoffs make it unfavorable for high-speed cache.

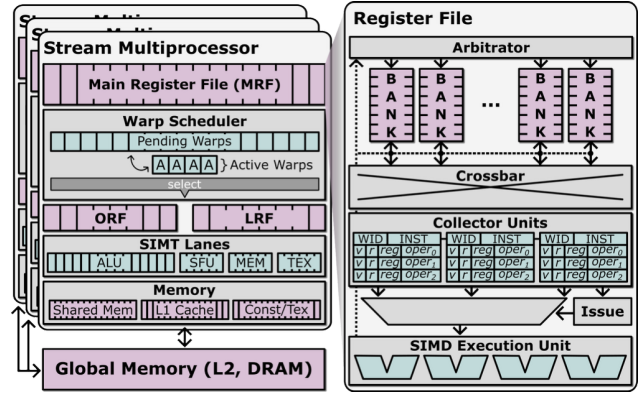

Fig 2: Organization of stream multiprocessor (SM) and reg file

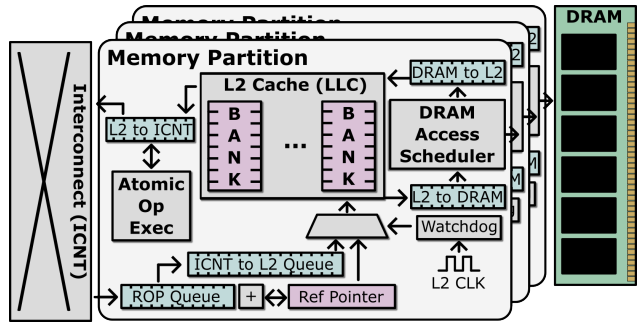

Fig 3: Organization of a memory partition, containing L2 slices, and additional refresh timing model incorporated in Accel-Sim

• We evaluate highly banked AOS L2 caches at iso-footprint in a modified Accel-Sim integrated on a verified NVIDIA Ampere RTX 3070 model. Our benchmarking reveals that AOS 1T1C integration can enable up to ~5.1× the performance per watt and ~6.1× memory density over a baseline HD-SRAM L2 cache.

## 2. Background
### 2.1. GPGPU Organization

Modern standalone GPUs comprise tens to hundreds of streaming multiprocessors (SMs), each hosting a programmer-controlled shared memory, private level 1 data (L1D) cache, register files (operand, last result, and main), and wide Single-Instruction-Multiple-Thread (SIMT) execution lanes containing arithmetic (ALU) and special-function units (SFU) (Fig. 2). When a kernel is launched, groups of threads in cooperative-thread arrays (CTA) or thread blocks are mapped to SMs, which are then partitioned into warps. Warps are time-multiplexed by the scheduler and executed in lock-step. To feed these wide SIMT lanes, each SM's register file contains tens of thousands of registers (~65k/SM in Blackwell), which are highly banked and, in smaller RFs, multi-ported (e.g., NVIDIA's Fermi architecture used 3-read/1-write (3R1W) in its operand register file (ORF) [7]).

Swaths of on-chip data are interleaved across LLC banks (slices), which are distributed across numerous memory partitions (Fig. 3). The memory partitions are connected to SMs through an on-chip interconnection network; each partition contains L2 slices, request schedulers, and a memory controller for off-chip DRAM (GDDR6X or HBM) [20]. In Ampere-class GPUs, there are eight partitions, each housing two L2 banks [64]. The L2 employs a write-back policy with respect to global memory [21].

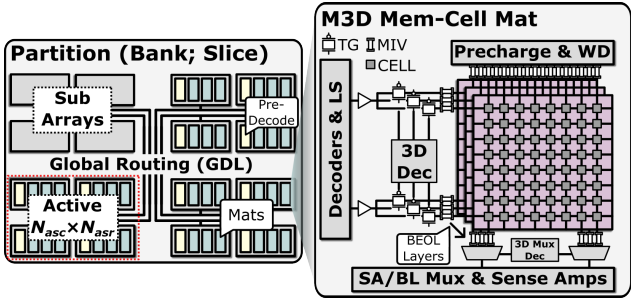

**Fig 4:** Organization of a monolithic-3D (M3D) bank/slice. Active subarray rows ($N_{asr}$) and columns ($N_{asc}$) often called sub-bank

### 2.2. Organization of a Memory Bank

Memory banks (slices) are topologically organized into a matrix of subarrays interconnected by global routes that comprise address, broadcast, and distributed data lanes [22]. These global routes, sometimes referred to as the global data line (GDL), often employ an H-Tree routing topology in RC-based memory simulators. When a transaction is received, it is routed over the GDL to a set of 'active' subarrays within each column ($N_{asc}$) and row ($N_{asr}$), which each deliver a sub-block of the aggregate block size of the bank. In some contexts, such as the nomenclature adopted by CACTI [23], this group of concurrently activated subarrays ($N_{asc} \times N_{asr}$) is referred to as a sub-bank. The number of transactions that can be issued to a bank per cycle is limited by the number of ports ($N_p$); however, with pipelining, it is possible to operate sub-banks in parallel if their activations are tracked, though they cannot be *issued* concurrently if $N_p = 1$. Each subarray is composed of a pre-decoder and a set of mats, which operate concurrently. Based on the operation concurrency, the maximum bandwidth ($BW_B$) for the bank with split read/write paths in a uniform cache access (UCA) model can be approximated as:

$$BW_B \approx N_p \times \max\left(\frac{W_{Block}}{t_{precharge}+t_{mat_{read}}}, \frac{W_{Block}}{t_{mat_{write}}}\right) \quad (1)$$

Where $W_{Block}$ is the block width per bank, and $t_{precharge}$ is the precharge latency. This dependence on mat latency makes cell access time a critical memory parameter, as discussed in §5.2. A mat contains an array of memory cells, and local peripherals used to drive data in and out of the array, typically decoders, pre-chargers, write drivers, sense-amplifiers (SAs), multiplexers (bitline, sense-amp), and level-shifters (when high-voltage swings are employed). We adopt a monolithic 3D (M3D) mat design in which each level is individually accessed using a 3D decoder that drives a set of transmission gates, allowing access to a specified level in the BEOL memory through the decoder/level-shifter drivers (Fig. 4). This 3D-decoded access scheme integrates a level-multiplexer and mux-decoder that allows BL sharing of FEOL SAs. In an M3D design, peripheral circuits are tucked under the memory array and connected by BEOL MIVs, which offer higher I/O density than TSVs [15].

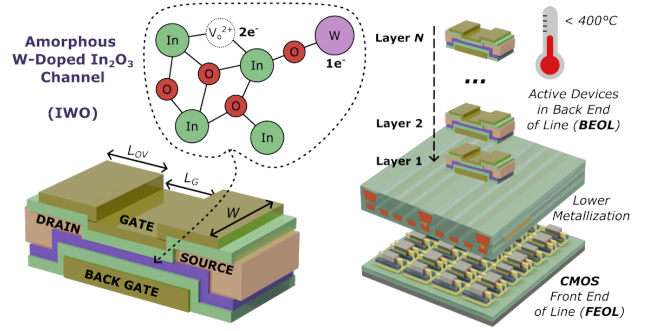

**Fig 5:** Double-gated IWO transistor structure and geometry, active monolithic 3D integration of devices above the FEOL

### 2.3. Oxide Semiconductor Transistors in the BEOL

Amorphous oxide semiconductors (AOS) such as indium oxide ($In_2O_3$) are a class of semiconducting oxides that exhibit moderate electron mobility (~20 cm$^2$/V·s), in which conduction is primarily governed by donor-like defects ($N_D$) such as oxygen vacancies. Dopants such as germanium (Ge), tin (Sn), or tungsten (W) are used when $In_2O_3$ is employed as a channel material to curb the formation of defects, thereby improving its stability and increasing its threshold voltage ($V_{th}$) [24]. Leading foundries (TSMC, Samsung) and research institutes (IMEC) are actively working on AOS channel materials for the integration of BEOL memories [25]-[27]. Demonstrations of stacked AOS transistors have been characterized in up to ten monolithic tiers and scaled to 10 nm gate length ($L_g$) [28]-[29].

### 3. Simulation Methodology

To build a cohesive evaluation of the design, technology, and system-level integration of BEOL-compatible AOS memories (Fig. 5), we employ a precise quantitative study that utilizes finite-element physical models, SPICE simulation, and cycle-accurate GPU simulation. Modeling of lab-measured double-gated (DG) long-channel W-doped $In_2O_3$ (IWO) transistors is performed in Sentaurus Technology CAD (TCAD), from which a scaled 7 nm technology model is developed ($L_g$ = 15 nm, $L_{ov}$ = 30 nm), and measured ($I_d$-$V_{ds}$, $I_d$-$V_{gs}$, and $C_{gg}/C_{gd}/C_{gs}$ parameters) for varying donor-defect densities ($N_D$). Extracted parameters are utilized to develop ML-assisted compact models [30] used in subsequent SPICE simulation, while Si-CMOS reference circuits are built with the ASAP7 PDK [81]. NS-Cache [17] is utilized to conduct an exhaustive search of the power, performance, and area (PPA) of various bank and subarray configurations presented in this work, using 7 nm FinFET predictive libraries. A baseline HD-SRAM cell of 0.0262 µm$^2$ is used based on the advanced foundry 7 nm platform technology [31]. Accel-Sim/GPGPU-Sim's verified NVIDIA Ampere RTX3070 model is used for system performance evaluation. We extend Accel-Sim's memory-partition timing to incorporate AOS refresh overheads, enabling cycle-accurate assessment of their impact on kernel execution (Fig. 3).

## 4. Scaling Register Files, CTAs, and Warps
### 4.1. On the Lifetime of Operands

The GPU register file must be able to deliver up to two source operands and one destination per thread for each warp-wide instruction within a single cycle (~1 ns in NVIDIA Volta to Hopper architectures) [32]. Consequently, the underlying register memories must (1) operate at high speed (~sub-nanosecond), and (2) be heavily banked or multi-ported to service multiple read/write requests concurrently. In some cases, (2) is taken to an extreme, as illustrated by the Intel Itanium microprocessor's 12R10W register file [33]. Capacitive memories that require a refresh operation seem ill-suited as a register memory, as refresh operations temporarily remove a bank from service, thereby hindering the high-speed requirement. Nevertheless, this has not impeded the proposal of eDRAM register memories in the literature, as seen in Si 3T1D [34] and FD-SOI 4T0C [35] cells. GainSight [36] argues that the key to understanding a persistent memory's suitability for a level in the memory hierarchy lies in the lifetime of data blocks (i.e., how long data is retained and utilized). Using Accel-Sim, we track the lifetime of registers in each SM, as defined by GainSight (Fig. 6a), on a set of randomly selected benchmarks from Rodinia [37]. We find that over 99% of register operands are overwritten or evicted within $10^5$ cycles (~100 µs on a 1 GHz SM clock). Although this may be beyond the retention of a gain cell on an Si or FD-SOI platform (~6 µs) [35], it is far below the achievable retention times in AOS gain cells, which enable retention times of (milli)seconds through low leakage.

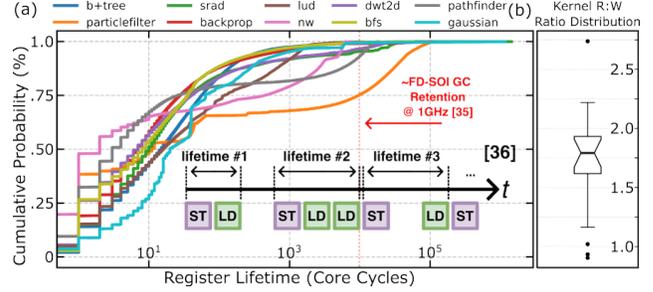

Fig 6: (a) Measured reg lifetimes on select Rodinia [37] benchmarks. Nearly all regs are limited to a $10^5$-cycle boundary (b) register read:write ratio distribution tracked per kernel

### 4.2. Alternative Paths to Multi-Porting

Multi-porting, which enables multiple accesses per cycle to a memory bank, is often employed in register files in both multi-core CPUs and GPUs. Multi-porting may be realized through several means: banks can be replicated [38], time-multiplexed on a faster clock domain [39], or virtually multi-ported through banking (at the cost of bank stalls during contention). The following study places particular emphasis on cell-level multi-porting, as (1) the speed of AOS memories is lower than Si (lower mobility), (2) the read and write paths in a gain cell are intrinsically bifurcated, and (3) leveraging M3D-stacking opens the opportunity for compact means of multi-porting layout. In SRAM, cell-level multi-porting (MP) can quickly become unfavorable due to interconnect congestion and increased transistor counts (~4T per write port and 2T per read port) that exponentially inflate cell size (~O($N_P^2$)) [40]; nevertheless, cell-level MP is still employed in CPUs with modest register counts to meet high speed single-cycle latency targets, (e.g., IBM Power & Intel Itanium [33],[41]). Because the GPU ORS employs additional read ports, and kernel register accesses are read-heavy (Fig. 6b), we focus on read multi-porting. Although additional write ports in a gain cell only require one transistor, the overhead of level shifting on write paths [17] imposes significant area penalties for duplicated periphery.

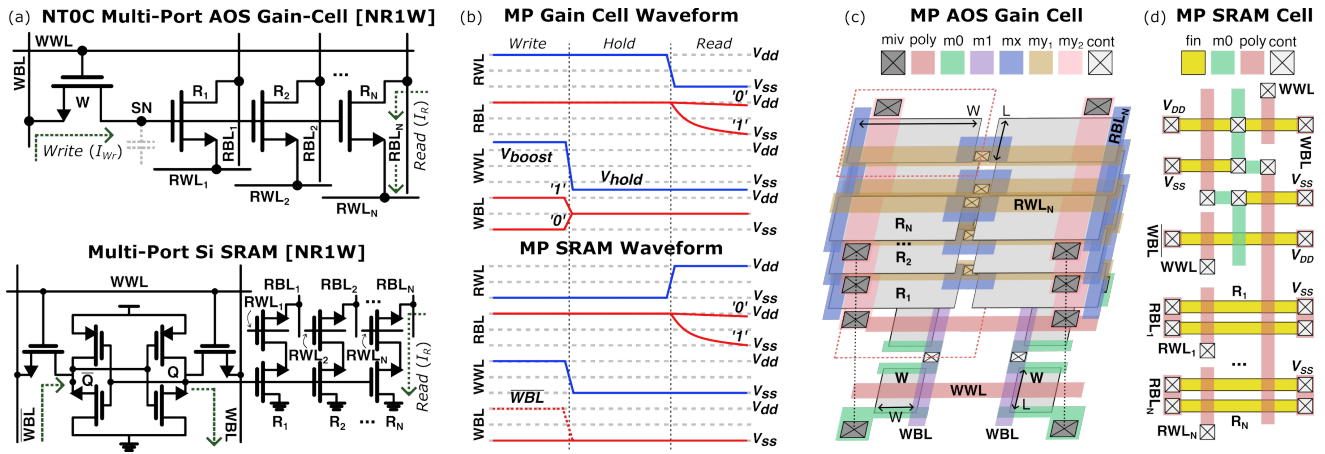

Fig 7: (a) Multi-ported AOS gain-cell (GC) and SRAM schematics, (b) cell operation diagrams, (c)-(d) physical layout and connectivity

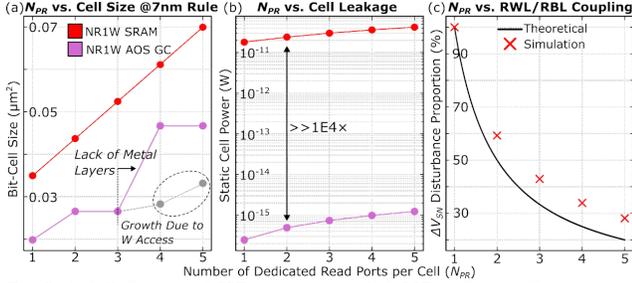
Fig 8: (a) Cell size of MP memories, (b) MP cell standby power, and (c) MP AOS RWL/RBL coupling, as a function of read ports

In Fig. 7a, we illustrate a conventional SRAM with split read (2T) and write paths (bidirectional), extrapolated from a 10T-derived SRAM design in [15], and our proposed AOS (DG-IWO) gain cell with NR1W ports. Fig. 7c and 7d illustrate the layout of each cell. For 7 nm BEOL design rules, we adopt the N7 metal-x ($M_x$, 40 nm) and metal-y ($M_y$, 76 nm) metallization pitches discussed in [42], an MIV pitch (60 nm) discussed in [43], as well as the CPP (54 nm) and fin pitch (27 nm) matching the foundry's 7 nm platform technology [31]. We assume the utilization of up to five $M_x$ and five $M_y$ layers. Because the number of stacked tiers in the AOS gain cell scales with the port count, Fig. 7c depicts abstract "functional" metal layers, and metallization limits are analyzed in further sections. Under these rules, we estimate that an 8T (1R1W) SRAM cell consumes 33.2% more area than a 6T SRAM cell, closely resembling the ~30% footprint increase observed in IBM's 65 nm PD-SOI process [44]. In both cells, the transconductance of a read transistor ($R_i$) is used to sink current from the pre-charged read-bitline (RBL) based on the stored value (SN in the gain cell, Q in the SRAM). For the gain cell, this operation requires applying a differential voltage to the $V_{ds}$ of the selected cell by driving the RWL to $V_{ss}$. In contrast, in SRAM, a pulse of $V_{dd}$ is applied to the gate of the read-gating ($R_G$) transistor, allowing the current to sink (Fig. 7b). A challenge posed by each read port in SRAM is that each read path adds additional subthreshold leakage from the precharged RBL, as a potential difference exists across each 2T read path. On the contrary, both the RWL and RBL (attached at the read port source and drain) are peripherally controlled in the 1T read path of the MP gain-cell; thus, leakage is suppressed, and static power can be derived from the cell retention:

$$P_{static} \approx \frac{C_{SN} \Delta V_{SN}^2}{t_{ret}}; \text{NT0C, 1T Read Port} \quad (2)$$

Where $C_{SN}$ is the storage node capacitance, $\Delta V_{SN}$ is the change in stored voltage before a refresh is issued, and $t_{ret}$ is the retention period. In [17], the trade-off between retention and access time is discussed in an IWO 2T0C using the $V_{th}$, and the hold/boost voltage ($V_{hold}$, $V_{boost}$), following which we study parameters of a cell with an optimized $R_i$ width ($W_{RA}$) of 150 nm, a write transistor ($W$) with a nominal width ($W_{WA}$) of 30 nm, aiming for a write-access time ($t_{wr}$) of 400 ps, $t_{ret}$ of 10 ms (~100× requirement in §4.1), $V_{hold}$ of -0.4 V and $V_{boost}$ of 1.2 V.

The $C_{gg}$ of $\sum R_i$ dominates $C_{SN}$, and $t_{wr}$ is inversely proportional to write current ($I_{wr}$), which scales with $W_{WA}$. Therefore, to preserve high write speed as read ports ($N_{PR}$) are added, $W_{WA}$ is widened in proportion to $N_{PR}$, increasing both cell size and static power (Fig. 8a and 8b). Additionally, since the pitch of the upper metallization is relaxed (~18× the pitch of $M_y$), cell stacking reaches a ceiling with $N_{PR}$ = 3, resulting in a sharp increase in footprint as two read transistors are integrated into each level. Nevertheless, we find that an NR1W AOS gain cell consumes over four orders of magnitude less standby power than SRAM, while occupying a smaller footprint.

An additional benefit of multi-porting AOS gain-cells can be elucidated from the theory presented for split-gated AOS transistors [45]. It is posited that the capacitive coupling phenomenon in AOS 2T0C gain cells, which refers to the modulation of the storage node voltage ($V_{SN}$) caused by pulsing the BL/WL voltage, is proportional to the ratio of all parasitically coupled capacitances to the storage node, imposed by the charge neutrality condition. This modulation reduces the sense margin and heightens read disturbance. However, as $N_{PR}$ is increased, so too is the number of capacitances coupled to the SN, leading to a theoretical 4× reduction in read capacitive coupling in 4R1W over 1R1W:

$$\Delta V_{SN} \propto \frac{W_{WA}}{W_{WA}+2\times W_{RA}\times N_P} \Delta V_{WWL} = \frac{W_{RA}}{W_{WA}+2\times W_{RA}\times N_P} \Delta V_{RWL} \quad (3)$$

In Fig. 8c, we plot the theoretical reduction in RBL/RWL coupling and the simulated coupling measured in SPICE as a function of the number of read ports. In simulation, the reduction measured in 4R1W is closer to 3×. Coupled with a split-gated AOS transistor geometry, the multi-porting of AOS gain cells provides an avenue for suppressing capacitive coupling without requiring high secondary gate bias voltages.

### 4.3. Macro-Level Performance, Power, and Area

Before evaluating the performance of NR1W cells at the bank level, we first examine the intrinsic cell-level limitations imposed by the single-transistor (1T) read port of the AOS gain cell, which places constraints on the allowable memory array dimensions. In the following, a subscript $s$ denotes selected lines/cells, and a subscript $u$ denotes unselected lines/cells in the same port. (1) *IR Drop:* Because the $RWL_s$ is used to sink current from $RBL_s$ when the SN stores a '1', the current must travel through the entire $RWL_s$ to the driver (sink), creating a voltage divider effect that reduces the $V_{gs}$ of $R_s$, thus limiting worst-case read speed as the number of columns ($N_{col}$) increases. (2) *Sneak Path:* During the read-out of a '1', the $RBL_s$ discharges over $R_s$ onto $RWL_s$; however, each $RWL_u$ is held at $V_{dd}$, thus leading to a reverse polarity $\Delta V_{ds}$ over each $R_u$, which in the worst case (i.e., when each SN stores '1') causes $RBL_s$ to prematurely settle before a sufficient read margin (RM) can develop (Fig. 9a), thereby limiting the number of rows ($N_{row}$). To quantify these effects in the upper limit, we simulate the read port of a 5R1W cell (for sizing) with an IR drop imposed by $N_{col}$ = 512 over varying $N_{row}$ and $V_t$, in the worst-case column data pattern of all '1's (Fig. 9b). Read margin is measured as the $V_{RBL}$ drop between '0' and '1' stored at the SN; we track (i) the peak RM, (ii) the time at which RM saturates and (iii) if/when the RM crosses the 200 mV level required for a 100 mV sense threshold. In the literature, the conventional guidance for architecting an AOS GC read port is to use a lower $V_t$ than the write port counterpart, since this will increase sink current for an increased $C_{RBL}$ (§5.2.1) [17], [46]-[47]. However, we refine this view: in cases with very few rows (i.e., $N_{ROW}$ = 64), a lower $V_t$ does accelerate read speed. However, as $N_{ROW}$ is increased, this relationship quickly becomes parabolic, and a higher $V_t$ is required to curb read failure caused by sneak path leveling. From this, we set a maximum $N_{ROW}$ of 64 (extendible to 128 if a folded BL layout is used [48]) as the upper boundary.

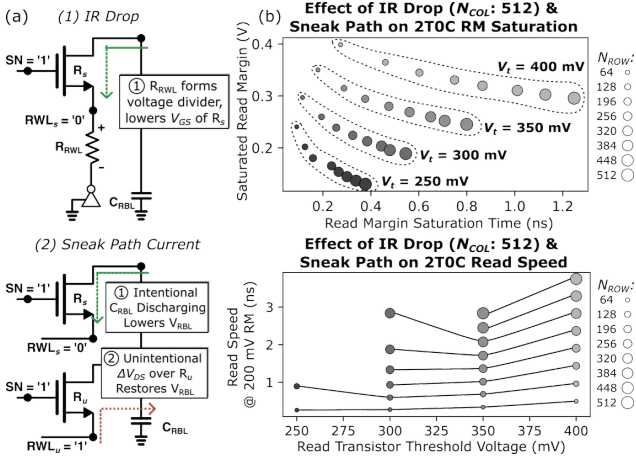

**Fig 9: (a) Illustration of IR drop and sneak path challenges using single AOS transistor port, (b) Effects on read margin and delay**

In NVIDIA's Fermi architecture, each MRF bank has a capacity of 8kB, each 16B wide with 32b per register [7]. We utilize the findings from the prior section in NS-Cache to model 8kB multi-ported SRAM and AOS gain-cell banks with a 128-bit $W_{block}$. We constrain the random cycle time (RCT) of each bank to 750 ps and exhaustively sweep design points with up to 8×8 subarrays per bank and mats per subarray, as well as up to 2 M3D integrated tiers of memory ($N_L$). The outcomes (area, static power, and dynamic energy consumption) as a function of $N_{PR}$ are plotted in Fig. 10a, for the minimum area 8 kB bank configuration under timing and sizing constraints. With a single BEOL tier, an AOS 2R1W bank can be placed in relatively the same footprint as an equivalent 1R1W 8T SRAM bank, and under the case that two tiers of memory can be integrated, a 3R1W AOS GC bank using cell-level multi-porting can integrated into the space if a 1R1W SRAM bank at ~0.76× in footprint without sacrificing RCT. Reductions in static power are not as striking as at the cell level, owing to the observation in [17] that increased capacitance of AOS devices necessitates larger drivers to optimize latency, thus increasing the leakage of peripheral circuits, however the static power can be dramatically reduced by 72.3%-79.1% over SRAM, significant considering the static power of register files is considered to consumer ~17% of SM static power in NVIDIA's Volta GPUs [49]. The key drawback, however, is that the dynamic energy consumption, especially for write operations in the AOS gain-cell, is often higher than that of SRAM, even when including the inter-bank-level routing power. This increase is attributed to the higher WWL switching energy ($\propto C_{WWL} \times \Delta V_{WWL}^2$) and constraints on the mat size, resulting in higher subarray activations and, consequently, higher mat-level access energy per operation (~5× that of SRAM with 1R1W). We observe that this access energy is reduced using stacking, as inactivated subarrays can be stacked on top of one another, leading to shorter routing. Prior work has reported that the high-access frequency of register files yields a ratio of ~6:1 in dynamic to static power consumption [50]. From this simplification, one can estimate that if total leakage is diminished, the increase in dynamic operation energy required to maintain the same overall power is ~16.6%, as the bank static power reduction approaches 100%. We plot this 'break-even point' in Fig. 10c alongside the dynamic energy consumption as a target, showing that in 60% of maximized density cases, the M3D AOS gain-cell bank maintains or improves overall register file power while improving footprint and portability.

To understand the implications on system performance, we utilize Accel-Sim's PTX mode to trace benchmarks that benefit from increased CTA size (i.e., warps per SM), increased SM count, and larger warp size (i.e., threads per warp). Each of these, respectively, is bound in some manner to the register file (i.e., by capacity, area, and bandwidth). For example, if a 3R1W device were adopted in place of an 8T SRAM, warp sizes of up to 96 threads (3× the NVIDIA Ampere baseline) may be enabled. If a 2R1W were used to halve the number of banks, the register file area per SM would be reduced by over 2×, allowing the integration of more SMs. Even if a simple 1R1W cell is adopted, the size advantages could be leveraged to increase

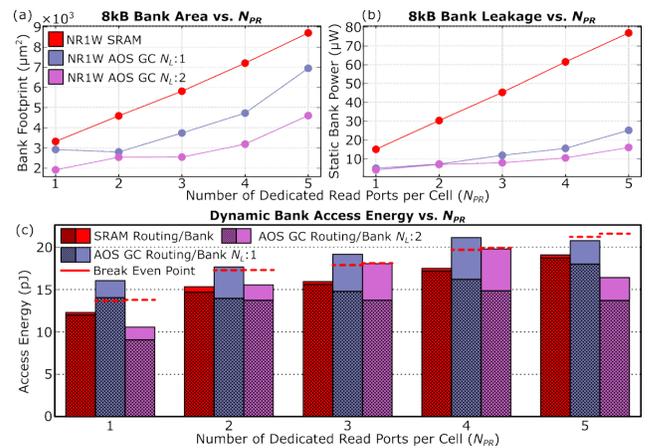

**Fig 10: (a) bank area comparison, (b) static power comparison, and (c) dynamic write access energy breakdown of MP memories**

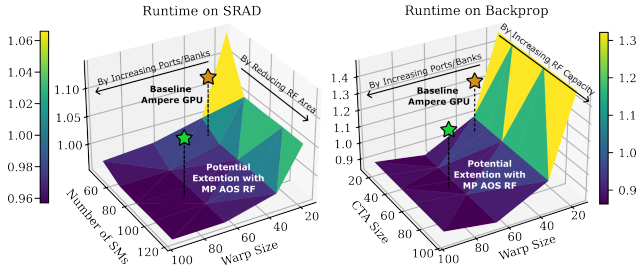

Fig 11: Warp, CTA, and SM scaling on compute-intensive workloads and translation using MP AOS GC integration normalized to NVIDIA Ampere RTX3070 baseline (Table II)

capacity, thereby doubling the CTA size. In Fig. 11, we demonstrate that arithmetic-intensive applications, such as backpropagation and srad from the Rodinia suite runtime, can be reduced by ∼10% by leveraging the density and portability of AOS gain-cell register files.

## 5. Scaling the LLC (L2)

The GPU LLC (i.e., the unified L2 in NVIDIA GPUs) reduces pressure on DRAM channels and hides long off-chip memory latencies that would otherwise stall SIMT execution pipelines [6]. Since the shift to post-Ampere architectures, the L2 cache has increased rapidly from 6MB to 120MB (Blackwell), primarily due to the data-hungry demands of data center applications, such as large language model (LLM) training and inference [51]. Nevertheless, one may (naively) assume that this points to the need for greater capacity at all costs; however, the relationship between workloads and the LLC is not so simple. Generally, memory-bound problems dependent on the L2 can be classified into one of three categories: (1) *Capacity-limited* problems, such as large matrix operations, where the kernel's working set exceeds the effective LLC capacity [52]. (2) *Bandwidth-limited* problems, such as database systems in which query-execution time is bound by the peak L2 bandwidth [53]. (3) *Latency-limited* workloads with irregular or fine-grained control flows, such as graph analytics, in which time to a hit is critical due to their data-dependent dynamic behavior [54].

From a designer's perspective of AOS LLCs, addressing each of these challenges involves leveraging the intrinsic spatial density of M3D design. Because the AOS LLC area, and thus capacity, is tightly bound to the area of peripherals, the largest array size that supports a target sub-bank RCT should be considered to minimize the number of duplicate mats per subarray and the ratio between FEOL and BEOL active footprints. It is worthwhile to target lower operating voltages and reductions in parasitic capacitance, as these qualities increase peripheral size due to bulky I/O level-shifting and large driver sizes [17]. As discussed in §2.2, the cache bandwidth ceiling is bound by the mat latency (and thus cell latency), the number of ports, and the number of banks. As will be discussed in the following subsections, the limitations of

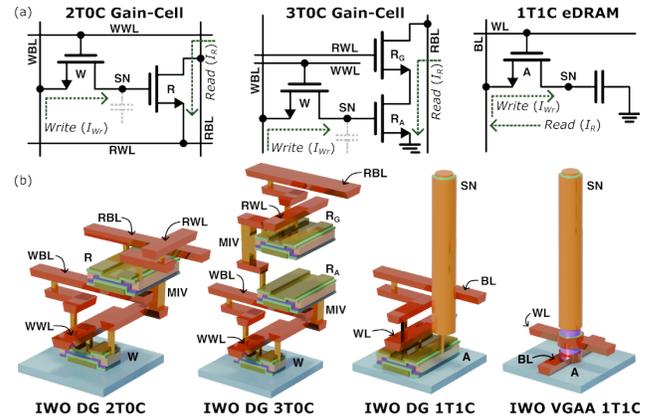

Fig 12: (a) AOS capacitive memory schematics and operating currents (b) 3D illustration of M3D integrated bit-cells, including double gated (DG) and vertical gate-all-around (VGAA) designs

AOS devices (i.e., lower current density and parasitic capacitance) may increase minimum $t_{mat}$, thus lowering the bandwidth ceiling per bank. Hence, adjusting to increasing the number of banks is critical to improve bandwidth, while understanding that: (1) cache fragmentation leads to increases in tag addressing widths, and thus tag capacity [55], (2) increasing banks inflates routing complexity, which, unless offset by smaller banks, dominates total L2 latency [56].

**Table I. AOS Cell Topology Assumptions in 7 nm Platform**

| Cell Topology | Contact | | Shared | Cell Size | Advantages | Limitations |
|---|---|---|---|---|---|---|
| 2T0C GC | Read | BL | N | 0.0195μm² | (+) Access Speed (+) Lower R Loading | (-) Small $C_{SN}$ (-) Sneak Path (-) IR Drop (-) Split R/W Periph. |
| | | WL | Y | | | |
| | Write | BL | Y | | | |
| | | WL | N | | | |
| 3T0C GC | Read | BL | Y | 0.0251μm² | (+) No Sneak Path (+) No IR Drop | (-) High Leakage (-) Larger Cell (-) Higher R Loading (-) Split R/W Periph. |
| | | WL | N | | | |
| | Write | BL | Y | | | |
| | | WL | N | | | |
| 1T1C eDRAM | BL | | Y | DG: 0.027μm² VGAA: 0.0182μm² | (+) Dedicated $C_{SN}$ (+) Fewer Periph. | (-) Destructive Read (-) Slower Access |
| | WL | | N | | | |

### 5.1. Comparison of AOS Cache Candidates

As discussed in depth as part of the constraints on array sizing in §4.3, the overall density and static power efficiency of AOS 2T0C integration are limited by a set of intrinsic (i.e., sneak path current, IR drop, small $C_{SN}$) and extrinsic (i.e., large buffers, level-shifter overhead, split-peripherals) factors. Thus, it is worthwhile to concurrently study alternative cell topologies using AOS devices that remedy some of the shortcomings of the 2T0C gain cell, while investigating the extent to which the relaxed timing requirements of the shared GPU LLC may mitigate the shortcomings of AOS 2T0C. We consider two alternative topologies for the following ablation studies in this work: (1) the 3T0C gain-cell, which adds an additional read-gating transistor ($R_G$) and shifts the RWL to the read control gate (the same read-operation principle as the 8T SRAM). (2) a BEOL-compatible 1T1C eDRAM, of which the access transistor (A) is an AOS device and incorporates a dedicated stacked capacitor as the $C_{SN}$ [13]. The M3D layouts and operational principles of the three proposed topologies are illustrated in Fig. 12. The contact sharing assumptions, cell sizes, and technological advantages and limitations are summarized in Table I. The remainder of this subsection discusses the tradeoffs of AOS 3T0C and 1T1C designs in terms of array sizing before proceeding to PPA comparisons at the macro level.

**5.2.1. Leakage and Speed in AOS 3T0C:** In §4.2, we briefly note that a shortcoming of the 2-transistor (2T) readout port in 8T SRAM is the increased static power of the cell, a function of the potential difference created by the pre-charged RBL and $V_{ss}$ connection over the read port. To achieve high speed in 8T SRAM (where differential BL sensing isn't used), the $V_t$ and fin count (thus, $W_{eff}$) of the read gating transistor ($R_G$) and the read access transistor ($R_A$) are modulated [44]. Though lowering the $V_t$ improves speed (higher $V_{ov}$), it increases the leakage ~exponentially in tandem. The schematic equivalence of the 2T read port in an AOS 3T0C lends itself to this same tradeoff, which we study using SPICE in the case of a single-fin CMOS 2T read port and an AOS 2T read port with a 150 nm $W_{RA}$ and $W_{RG}$ (Fig. 13). We observe that the AOS 3T0C cell pays a higher price for high speed than its Si counterpart primarily due to (1) the reduced current density of the AOS channel transistor and (2) the higher parasitic capacitance created by large overlap regions, which increases the total bitline charge and thus the sink current density requirement of the read port for the same read speed. Therefore, to achieve sub-nanosecond RM development, an AOS 3T0C array with an $N_{ROW}$ exceeding 128 imposes cell standby power orders of magnitude higher than HD-SRAM (~14 pW). This, alongside the increased peripheral leakage (drivers) and cell-size disadvantages when compared to 2T0C and 1T1C, makes the DG IWO 3T0C an unsuitable choice for high-bandwidth, energy-efficient last-level cache memory (§5.2.3). Alternative AOS device geometries, such as the self-aligned gated structure with plasma-treated, conductive source/drain [62], may prove better suited for 3T0C integration; reductions in the BL capacitance contribution from device parasitics shift the read/leakage tradeoff downwards, showing a larger leakage suppression under the same read speed over Si (Fig. 13).

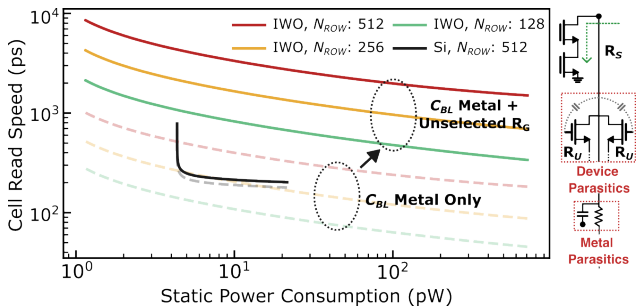

Fig 13: Read speed vs. leakage in 2T read port of 7 nm Si and AOS

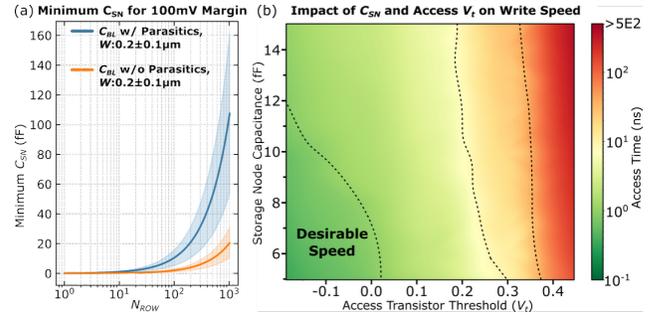

Fig 14: (a) Impact of cell parasitics on minimum $C_{SN}$ requirement, (b) relationship of $C_{SN}$ and access $V_t$ on write speed in AOS 1T1C

**5.2.2. Access Speed and Readout in AOS 1T1C:** Given the BEOL process compatibility of stacked capacitor fabrication, it is worthwhile to consider the advantages of a fully BEOL-compatible 1T1C eDRAM utilizing an AOS access transistor (Fig. 12a), reminiscent of that recently demonstrated by TSMC [13]. For one, split read/write paths in gain-cell topologies require additional peripherals to decode and drive both R/W paths, the footprint of which is truncated in an AOS 1T1C array using a single BL/WL per cell, thus delivering greater density and static power benefits. Additionally, using a dedicated capacitor, as opposed to solely the parasitic $R_A$ gate capacitance, allows us to raise $C_{SN}$ and thus enable more flexibility in retention, as higher cumulative charge is stored. However, these transformations also have limitations; for one, the $C_{SN}$ cannot be arbitrarily set, as the RM ($\Delta V_{BL}$) is a function of the BL capacitance ($C_{BL}$), SN capacitance ($C_{SN}$), and minimum SN voltage after retention losses ($V_{min}$):

$$\Delta V_{BL} = \frac{1}{1+C_{BL}/C_{SN}} \times (\frac{1}{2}V_{dd} - \frac{I_{leak} \cdot t_{ret}}{C_{SN}}) \quad (4)$$

The higher bitline capacitance contribution of AOS device integration increases the minimum $C_{SN}$ for 100 mV RM (Fig. 15a). We find that at $N_{ROW}$ = 128, the minimum $C_{SN}$ must be ~10 fF when the access device width is 200 nm ($V_{min}$ = 600 mV), and that the minimum $C_{SN}$ is a strong function of the access device width. The secondary bottleneck to consider is the decreased current density of the AOS access device, which increases access time. Since the readout of the 1T1C cell is destructive, a write-back must be issued during every read transaction; thus, access speed is a critical bottleneck in 1T1C arrays. If the RM for a desired $N_{ROW}$ sets the lower bound for $C_{SN}$, then the requisite RCT/access speed sets the upper bound. We study the relationship between access speed, $C_{SN}$, and $V_t$ in a 100 nm width IWO transistor using a $V_{cc}$ of 750 mV (Fig. 12b). To achieve sub-nanosecond cell access speed, it is desirable to use a smaller $C_{SN}$ (< 8-12 fF) with a low or negative $V_t$ (≤ 0 V), although the former comes at the expense of limiting $N_{ROW}$.

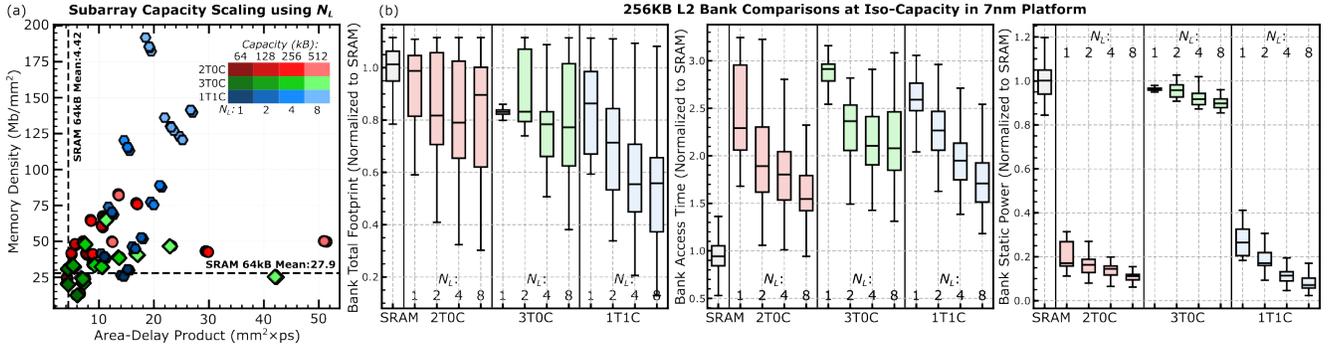

Fig. 15: (a) 3D stacking speed-density tradeoff demonstrated in a 64kB (baseline) subarray with 1 ns RCT (excluding write-back). (b) 256kB memory bank footprint, access time, and static power distribution comparison of AOS cache candidates and SRAM.

Based on the previous results, the following sections set $C_{SN}$ to 10 fF, an access device width of 300 nm (limiting $N_{ROW}$ to 64), and a $V_{hold}$ of -300 mV (requiring negative level shifting). Shown in Table I and Fig. 12b, a wide access device in the quasi-planar double-gated geometry imposes a restrictive cell footprint. Therefore, a vertical gate-all-around (VGAA) structure with a cylindrical channel is adopted to improve cell density [13].

**5.2.3. Subarray and Bank-Level Comparison:** To draw comparisons between macro-level implementations of AOS-based cache memories, we conduct two distribution-based studies: (1) given a single subarray with a variable number of mats and 1 ns RCT constraint, what is the effect on density and speed if multiple stacked memory layers are used to *increase capacity*? (2) Given a 256 kB bank with a variable number of sub-banks and nanosecond sub-bank RCT, what is the effect on area, speed, and static power if stacked memory layers are used to *increase density* (Fig. 13)? Given the lower access speed of 1T1C, the RCT restriction is placed on the read operation without the write back processes.

We discern from Fig. 15a, which illustrates the subarray study (1), that AOS cache memories exhibit a higher area-delay product than their SRAM counterparts, as expected due to their slower operation speeds resulting from the lower electron mobility in AOS devices and the differential sensing achieved using feedback in SRAM. The density benefits of gain-cell topologies taper off quickly as multiple tiers are integrated, trading off quickly with delay as the $N_L$ is increased, resulting in a reduced slope. One reason for this is the overhead of 3D decoding in gain cell topologies compared to 1T1C: aside from the additional peripherals (drivers) for split R/W paths, the WWL drivers in the gain-cell topology are larger to handle the high voltage swing requirement ($V_{hold}$ to $V_{boost}$), and thus the transmission gate used to decode the selected layer is proportionally sized to last stage of the driver, leading to a higher footprint cost per layer than the 1T1C counterpart with reduced WL swing. As a result, at eight layers we observe that a 512kB 2T0C subarray under a 1 ns RCT restriction can achieve memory densities of 82.87 Mb/mm$^2$ (~2.72× SRAM peak), 3T0C can achieve 65.68 Mb/mm$^2$ (~2.16× SRAM peak), and 1T1C can achieve 191.8 Mb/mm$^2$ (~6.1× SRAM peak).

To impose a limitation on the total footprint in our bank-level study, we estimate the area of a 256 kB read-optimized bank, which is reflective of the L2 partition capacity used in an NVIDIA Ampere GPU [20],[64], yielding a footprint of ~80,000 μm$^2$. We plot the distributions of bank footprint, access time and static power for AOS cache candidates in ascending $N_L$ in Fig. 15b. We first observe that switching the loading of the read decode path from the source of $R_A$ to the gate of $R_G$ increases partitioning of arrays, and thus larger footprints and bank latencies in 3T0C compared to 2T0C and 1T1C, that cannot be explained by cell size increases alone. Although not at the level of SRAM, the AOS 2T0C bank speed median and minimum values strongly outclass those of AOS 3T0C and 1T1C. Gain-cell topologies exhibit larger footprints than eDRAM and stronger leveling off of footprint reduction as a function of $N_L$. Similar to the prior subarray result, 1T1C achieves the highest footprint reduction as a function of the number of tiers among the three candidates. The reduced cell-level leakage significantly improves the overall consumption of the array in 2T0C and 1T1C, although the higher driver leakage (as a function of sizing) maintains averages in the ~10-40% range of SRAM, with substantial reductions realized using stacked arrays. Conversely, to maintain reasonable read speeds in 3T0C, the cell-level leakage is orders of magnitude higher than its 2T0C and 1T1C counterparts. At $V_t$ = 250 mV, the per-cell 3T0C static power is ~2.67 pW, resulting in bank-level static power consumptions comparable to that of SRAM. For the stated density, static power, and speed limitations, we exclude 3T0C from the following system-level benchmarking study, which investigates the implications of M3D AOS memory integration for enhancing LLC capacity and improving GPU performance.

## 6. Benchmarking Methodology

We evaluate the proposed integration of IWO 2T0C and 1T1C L2 caches using the cycle-accurate GPGPU simulator Accel-Sim [19]. We model the baseline system after a verified NVIDIA Ampere RTX 3070 GPU model [20], with system simulation parameters listed in Table II. Though the capacity of the L2 is small in Ampere GPUs (4 MB) compared to the state of the art (Blackwell, 126 MB), studying the implications on a validated, compact model can provide broader insights into the performance implications on larger GPGPU LLCs. To provide a comprehensive evaluation of performance, we randomly select 15 compute- and memory-bound applications from the Rodinia [37], Polybench [57], and DeepBench [58] benchmarking suites, targeting relevant workloads in scientific simulation, image/signal processing, graph analytics, linear algebra, and deep learning primitives (Table III). To achieve cycle-accurate replay in Accel-Sim, we utilize Accel-Sim's trace-driven mode, which employs dynamic SASS instructions logged using NVIDIA's NVBit instrumentation framework [59]. We modify the integrated memory partition model to account for the impact of (distributed) refresh operations on performance, which are incorporated into reservation failure statistics and discussed further at the end of §7.

**Table II. Baseline GPU Benchmarking Configuration**

| Parameter | Configuration |
|---|---|
| Number of SMs | 46 |
| Schedulers per Core | 4, Loose Round Robin [60] |
| GPU Memory Interface | 256-bit GDDR6 |
| GPU Memory Capacity | 8 GB |
| L1/Shared Memory Capacity | 128 KB per SM |
| Register File Capacity | 256 KB per SM |

**Table III. Evaluated Benchmarks and Domains**

| Application | Abbrev. | Domain |
|---|---|---|
| Covariance Computation [57] | cov | Pattern Recognition |
| Particle Filter [37] | pfil | Medical Imaging |
| Discrete 2D Wavelet Transform [37] | dwt2d | Image/Video Compression |
| Convolutional Neural Network Training [58] | cnn | Deep Learning |
| Matrix Transpose and Vector Multiplication [57] | atax | Linear Algebra |
| Back Propagation [37] | backprop | Pattern Recognition |
| Matrix Vector Product and Transpose [57] | mvt | Linear Algebra |
| Pathfinder [37] | pfin | Dynamic Programming |
| 3D Convolution [57] | 3dconv | Image Processing |
| GEMM Kernel Inference [58] | gemm | Deep Learning |
| Needleman-Wunsch [37] | nw | Bioinformatics |
| B+ Tree [37] | b+tree | Search |
| RNN + GRU Training [58] | rnn | Deep Learning |
| Correlation Computation [57] | corr | Signal Processing |
| 3 Matrix Multiplication [57] | 3mm | Linear Algebra |

To understand the impact of both high bandwidth and high capacity in scaled L2 AOS-based LLCs, we model four M3D-based L2 systems, two for each AOS memory type: (1) *Iso-Banking (IB)*: assuming the same footprint and number of data banks, what is the performance of an L2 that leverages stacking to maximize per-bank capacity? (2) *Iso-Bank Capacity (IBC)*: assuming the same bank capacity and aggregate footprint, what is the performance of an L2 that leverages stacking to decrease bank size and maximize the number of banks? The parameters of each evaluated system are shown in Table IV. We set a relaxed L2 footprint constraint per memory partition of 200,000 μm². In *IB* studies, we halt increasing $N_L$ once no configurations exist within the footprint constraint. The L2 clock domain is used to model the operating frequency, and the raster operations pipeline (ROP) latency is used to model the L2 latency [54]. Because memory partitions (and thus L2 banks) are highly distributed using complex networks on chip (NoCs), we only consider alterations to the ROP latency based on local changes in latency within each partition relative to the SRAM baseline. Additionally, the overhead of tag memories is omitted from this study; however, we note that tag/directory overhead will increase proportionally to capacity [17] and with additional partitioning [55].

**Table IV. Benchmarking L2 Cache Parameters**

| Config | Memory Type | # Banks | L2 Capacity | $N_L$ | L2 Clock Domain | L2 Area/ Partition | ROP Latency | Ref Period |
|---|---|---|---|---|---|---|---|---|
| Baseline | SRAM | 8×2 | 4 MB | 1 | 1132 MHz | 160,328 μm² | 187 cyc. | N/A |
| 2T0C IB | IWO 2T0C | 8×2 | 8 MB | 2 | 1067 MHz | 130,453 μm² | 184 cyc. | 859 μs |
| 2T0C IBC | IWO 2T0C | 8×8 | 16 MB | 8 | 1132 MHz | 195,044 μm² | 188 cyc. | 215 μs |
| 1T1C IB | IWO 1T1C | 8×2 | 32 MB | 8 | 724 MHz | 175,771 μm² | 190 cyc. | 244 μs |
| 1T1C IBC | IWO 1T1C | 8×16 | 32 MB | 8 | 724 MHz | 166785 μm² | 190 cyc. | 244 μs |

## 7. Evaluation

In Fig. 16, we track total performance (instructions per cycle, IPC), application runtime, and performance per watt for each benchmark and its geometric mean. Based on pairings where performance is optimized according to *IBC* configurations or high-capacity 1T1C configurations, benchmarks can be broadly categorized as either *bandwidth-limited* (e.g., gemm) or *capacity-limited* (e.g., correlation), respectively. Both 1T1C-*IB* and 1T1C-*IBC* deliver elevated mean performance over the SRAM baseline, indicating that the elevated capacity made possible using AOS 1T1C integration surpasses the limitation on maximum per-bank bandwidth. However, there is a clear distinction between 1T1C-*IB* mean IPC and runtime, and the IPC and runtime in the lower-capacity 2T0C-*IBC*. This reinforces the necessity for bandwidth in both bandwidth-limited applications and the maintenance of performance in compute-bound applications. 1T1C-*IBC* demonstrates the highest performance, yielding a ∼8% increase in IPC over SRAM and ∼2.7× mean increase in Perf/W. In select cases, IPC is up to ∼38% higher than the baseline, and Perf/W can be more than 5× higher. Although a few cases suffer slightly from the lower frequency in 1T1C-*IBC* cases (e.g., particlefilter), these performance reductions are relatively negligible (< 1%). Although 1T1C-*IBC* demonstrates the highest mean Perf/W, the highest peak Perf/W is observed in 2T0C-IB cases such as gemm, and rnn. This should not be taken at face value: because power is the energy consumed over time, and the runtime of applications is longer in 2T0C-*IB* cases than in 1T1C-*IBC*, the higher performance per watt does not always indicate greater efficiency. Based on the performance gains observed in both IBC cases, we determine that the best use case (in terms of performance) utilizing AOS memories is leveraging stacking to reduce bank size and increase bank count.

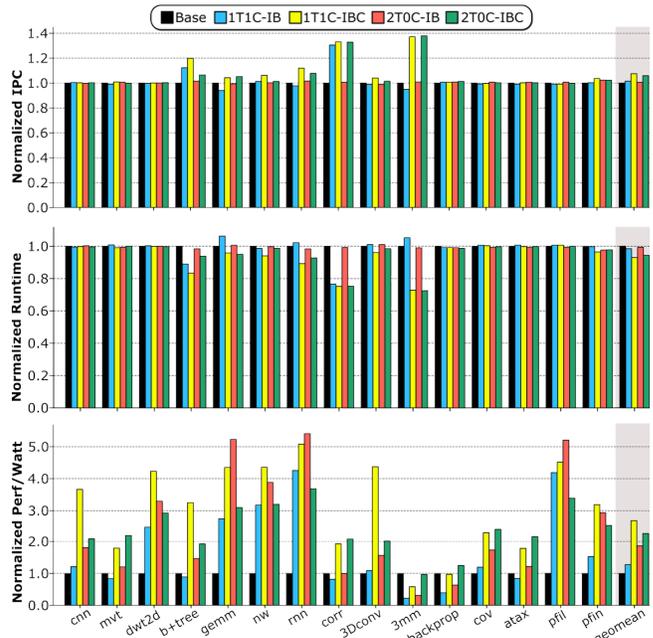

Fig 16: Instructions per cycle (IPC), runtime and performance per watt comparisons of evaluated systems, normalized to baseline

We present miss rates and energy consumption breakdowns per benchmark at the bank level in Fig. 17. This analysis yields two key findings. First, strongly reduced miss rates do not necessarily equate to higher performance, even when access frequency is high. Take, for example, cnn, which has the highest read access frequency among all benchmarks and a ∼40% reduction in miss rate in 1T1C cases but only demonstrates a ∼2% increase in IPC due to its ultimately compute-bound nature [61]. However, a goal of the GPGPU L2 is to alleviate the bandwidth requirements on the off-chip DRAM and reduce costly off-chip accesses, to which the corresponding reductions in miss rate aid in overall energy reduction. Second, at the bank level, SRAM leakage is often the dominant consumer of energy in the L2 cache; however, in cases with many read and write accesses (e.g., 3mm), dynamic energy consumption can quickly outpace static power. As was seen in §2, array-level dynamic power consumption is increased in AOS caches due to both increased capacitances and voltage swings, and restrictions on array size (which result in higher activation per transaction). This means that bank-level energy is dramatically increased in cases such as backprop and 3mm, which reduces the performance per watt. Reinforcing the prior statements on stacking methodology made in the preceding paragraph, we observe that *IBC* configurations consume less dynamic energy than their *IB* counterparts, as the size of each tiered array can be reduced, thereby imposing lower access energy. We also observe that infrequent refresh operations incur minimal energy costs across all AOS LLC configurations, with a breakdown shown for the case with the highest geometric mean refresh energy consumption: 1T1C-*IBC*, where line refreshes consume ∼0.8% of all energy (Fig. 17).

To understand the impact of extended retention (or infrequent refresh), we plot the distribution of modes of L2 reservation failures as a ratio of total L2 accesses per kernel across all benchmarks for *IBC* cases (Fig. 18). It is worth defining the most common non-refresh-based modes of reservation failures for context. A *line allocation failure* refers to when a new tag entry for the access is needed, but all ways in the target set are reserved by outstanding misses or pending fills. An *MSHR entry failure* occurs when a fresh miss status holding register (MSHR) entry needs to be allocated, but the table is full. An *MSHR merge failure* occurs when a miss for an address is already being tracked, but the per-MSHR merge list is full, and another requester cannot be recorded. Finally, a *miss queue failure* occurs when the miss queue can no longer allocate more incoming requests to off-chip DRAM. In our benchmarking, distributed refreshes occur for one cycle (1T1C) to two cycles (2T0C) in the L2 clock domain at a period of $t_{ret}/N_{ROW}/N_L$, as each layer in the 3D decoding scheme operates as an independent but peripherally coalesced matrix.

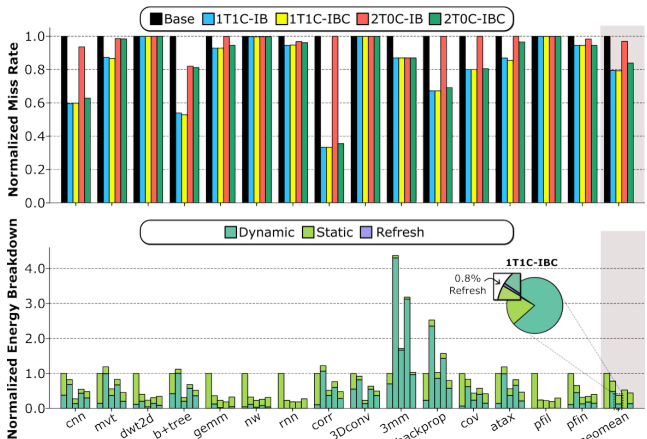

**Fig 17: Miss rate and corresponding system energy breakdown normalized to baseline. Refresh contributes negligible overhead.**

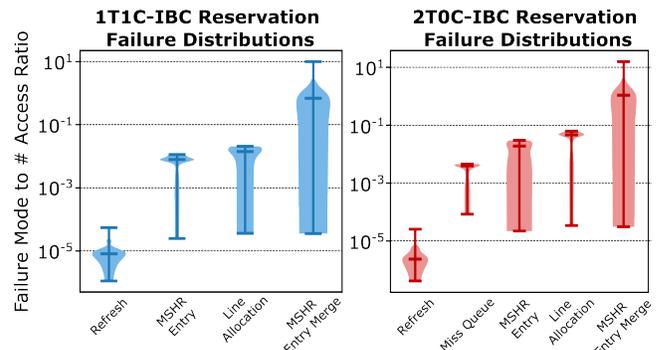

**Fig 18: Distribution of reservation failure modes in 1T1C- and 2T0C-IBC configurations gathered across benchmarked kernels.**

Notably, the peak frequency of refresh-induced read and write reservation failures at this periodicity is comparable to that of the lower bound of MSHR entry failures and is typically overshadowed by MSHR entry and line allocation failure modes by several orders of magnitude, indicating that retention levels in the tens to hundreds of ms range are sufficient for large GPGPU cache performance without incurring notable performance degradation due to L2 stalling. We also observe that the move from 4× to 8× capacity eliminates all miss queue failures in the 1T1C case, but heightens the distribution of line allocation failures, as contention over oversubscribed indices in each set (due to the higher hit rate) grows. Consequently, this notion poses the need to increase associativity alongside the set size [63].

## 8. Related Work

Several emerging memories have been the subject of study for GPU register file (RF) integration due to their density and static power advantages. Magnetic tunnel junction (MTJ) based multi-port racetrack memory [65], STT-MRAM [66]-[68], and SOT-MRAM [69] register files have been the focus of prior GPU system-level studies. Much of this body of work is centered on minimizing the shift-operation and write-speed bottlenecks of emerging non-volatile memory register files, which carry latencies on the order of ~tens of nanoseconds. Volatile Si eDRAM variants [34] and SRAM-DRAM hybrid memory [70] have also been studied at the system level as a compact RF solution. Low-leakage multi-ported SRAM designs [71] and FDSOI eDRAM [35] have been studied with GPU-register-centric integration in mind at the circuit and device levels.

STT-MRAM [72]-[73], Spintronic tape memories [74], and eDRAM [75] have also been the focus of GPU LLC expansion in prior work, typically benchmarked on Fermi architecture lines. Like their register file study counterparts, these works investigate microarchitectural methods for hiding long write latencies, minimizing the overhead of shift operations, and reducing the refresh overheads inherent to each memory type. Morpheus [5] discusses utilizing idle SM resources (L1D, Shared Memory, RF) as configurable extensions to the LLC.

Monolithic 3D integration has been studied in GPU memory and network subsystems, such as multi-tier register file banks and mesh networks [76] and the private L1 data cache [77]. Additionally, an M3D integrated NoC for GPU cache bypassing has been explored [78]. We note that these works are not technology aware and presume that multi-tiered CMOS integration is achieved with comparable performance to FEOL devices; therefore, they are opportunistic.

Finally, dense AOS-based memory systems have been the subject of a few benchmarking studies in CPU, TPU, and non-Von Neumann computing systems. [15]-[16] benchmark DCIM integration of hybrid 2T0C-RRAM and 2R1W AOS gain cells. [79] performs a benchmark of a Sn-doped $In_2O_3$ (ITO) 2T0C L1D extension in a TPU benchmark; however, the device-centric nature of the work leaves little room for discussion on macro- or system-level design assumptions and fundamental bottlenecks. [80] Focuses on the integration of IWO 2T0C as a TPU buffer memory, comparing it to other mainstream and emerging memories. The CPU cache-based IWO 2T0C study, as discussed in [17], examines system implications using GEM5. Although the study examines limitations on bank-level write, leakage, and retention implications in write design, there is limited discussion on read implications, and it lacks a study of multiple system configurations.

## 9. Conclusion

This paper presents a study on the integration opportunities of monolithically 3D stacked amorphous oxide semiconductor (AOS) memories in capacitive memory topologies to tackle GPU memory-system bottlenecks. By observation of the relatively short lifetime of register operands, we develop integration methods for a high-speed multi-ported AOS gain cell capable of delivering three times the read ports, roughly three-quarters of the bank size of comparable 1R1W SRAM. Furthermore, we investigate the integration of stacked BEOL-compatible 3T0C, 2T0C, and 1T1C memories, demonstrating that 2T0C can achieve densities 2.72× higher than SRAM without sacrificing the maximum operating frequency, and 1T1C can deliver up to 6.1× the density of SRAM in 8-tier configurations. Benchmarking on a baseline NVIDIA Ampere GPU in a modified version of Accel-Sim indicates that AOS-based 1T1C last-level caches can boost IPC by a geometric mean of 8%, and as high as 38%, and performance per watt up to 5.1× over the baseline HD-SRAM system. Such CMOS+X stacking, therefore, offers a manufacturable path to reclaim area, bandwidth, and energy headroom that conventional SRAM scaling can no longer provide, enabling larger warp sizes, higher SM counts, and/or denser LLCs without increasing die size.